\documentclass[pra,aps,twocolumn,epsfig,superscriptaddress,showpacs]{revtex4}
\usepackage{amsmath}
\usepackage{amssymb}
\usepackage{graphicx}
\usepackage{dcolumn}
\usepackage{bm}
\usepackage[colorlinks=false,dvipdfm]{hyperref} 
\usepackage{setspace}
\usepackage{amsfonts}
\usepackage{rotating}
\usepackage{makeidx}
\usepackage{amsthm}
\usepackage{amsmath}

\begin{document}

\title{Discriminating bipartite mixed states by local operations \footnote{Phys. Rev. A 101, 032316 (2020)}}

\author{Jin-Hua Zhang}
\affiliation{School of Mathematical Sciences, Capital Normal
University, Beijing 100048, China} \affiliation{Department of
Physics, Xinzhou Teacher's University, Xinzhou 034000, China}

\author{Fu-Lin Zhang}
\email[Corresponding author: ]{flzhang@tju.edu.cn}
\affiliation{Department of Physics, School of Science, Tianjin
University, Tianjin 300072, China}

\author{Zhi-Xi Wang}
\affiliation{School of Mathematical Sciences, Capital Normal
University, Beijing 100048, China}

\author{Le-Min Lai}
\affiliation{School of Mathematical Sciences, Capital Normal
University, Beijing 100048, China}

\author{Shao-Ming Fei}
\email[Corresponding author: ]{feishm@cnu.edu.cn}
\affiliation{School of Mathematical Sciences, Capital Normal
University, Beijing 100048, China}
\affiliation{Max-Planck-Institute for Mathematics in the Sciences,
D-04103 Leipzig, Germany}



\begin{abstract}

Unambiguous state discrimination of two mixed bipartite states via
local operations and classical communications (LOCC) is studied
and compared with the result of a scheme realized via global
measurement. We show that the success probability of a global scheme
for mixed-state discrimination can be achieved perfectly by the local
scheme. In addition, we simulate this discrimination via a pair of
pure entangled bipartite states. This simulation is
perfect for local rather than global schemes due to the existence of
entanglement and global coherence in the pure states. We also
prove that LOCC protocol and the sequential state discrimination (SSD) can be interpreted in a unified view. We then hybridize the LOCC
protocol with three protocols (SSD, reproducing and
broadcasting) relying on classical communications.
Such hybridizations extend the gaps between the optimal success probability of global and local schemes, which can be eliminated only for the SSD rather than the other two protocols.

\end{abstract}



\maketitle






\section{Introduction} \label{intro}
Since useful quantum information is encoded in quantum states, state
discrimination is one of the most crucial research topics in
quantum information processing \cite{Enk2002PRA}. In particular,
the unambiguous discrimination among
linearly independent nonorthogonal quantum states is of fundamental
significance in quantum information theory
\cite{Ivanovic1987PLA,Peres1988PLA,Dies1988PLA,Bennett1992PRL,Bergou2003PRL,Pang2009PRA}.
For the simplest state discrimination, one prepares a qubit in one of two known
nonorthogonal states, $|\Psi_1\rangle$ and $|\Psi_2\rangle$, and
sends it to an observer Alice. Alice's task is to determine the
state she received by positive operator-valued measure (POVM). The measurement
gives rise to three possible outcomes, $|\Psi_1\rangle$, $|\Psi_2\rangle$, and
\emph{inconclusive}, in which the last one is the price for perfect
discrimination. Such unambiguous state discriminations play an
important role in quantum key distribution \cite{Bergou2003PRL}
and the study of quantum correlations \cite{Roa2011PRL,Li2012PRA,Zhang2013SR}.

For multipartite quantum states, the
measurement strategies can be classified into two types: global
and local. The authors in
Refs. \cite{Chen2001PRA,Chen2002PRA,Ji2005PRA} investigated
nonorthogonal bipartite pure states discriminated via local
operations. The observer Alice applies the discrimination operation
on the first particle first. If she succeeds, the procedure ends.
Otherwise, she sends the state to the next
observer, Bob. Bob takes his discrimination operation on another
particle. It is found that there exist protocols whose optimal
state discrimination of local operations and classical communications (LOCC) are as good as global schemes.

To find out the essential role played by local schemes in state discrimination,
we construct a pair of
mixed bipartite states comprising two orthogonal vectors mixed with each other via classical probabilities, which contain no entanglement or global coherence.
These mixed-state discrimination problems have given rise to many novel outcomes by global scheme in Ref. \cite{Namkung2017PRA}. It is found that the optimal successful probability of global
mixed state discrimination can also be achieved perfectly by the
local scheme, which is observed in pure state cases \cite{Chen2001PRA}.

In order to see the essential difference in
identifying pure and mixed states, we can simulate the above-mentioned (separable) mixed-state discrimination by an entangled and globally coherent state. We find that this
simulation is bound to be perfect for local scheme, since local
POVMs eliminate the entanglement and global coherence that are
critical recourses encoded in the pure states. Thus, the pure-state
scheme does not necessarily show superiority to mixed ones. For the global
scheme, successful simulation only occurs for a few special cases. Generally, the mixed-state protocol is inferior to the pure-entangled-state one.

Another scheme is the sequential unambiguous state discrimination
(SSD) originated from one of the theories to extract information
from a quantum system by multiple observers
\cite{Bergou2013PRL,Nagali2012SR,Filip2011PRA} and put forward by
Bergou \emph{et al.} in Ref. \cite{Bergou2013PRL}. It is shown in Refs. \cite{Bergou2013PRL,Namkung2018SR} that SSD is
useful in quantum communication
schemes (e.g., the B92 quantum cryptography
protocol \cite{Bennett1992PRL}). The optimal
success probability of SSD was provided analytically in
Ref. \cite{Bergou2013PRL} and demonstrated experimentally
\cite{Solis-Prosser2016PRA}. Further investigations on
the optimized success probability of SSD with global measurements
are reported for both the pure states \cite{Pang2013PRA,Zhang2017arXiv,Namkung2018SR}
and mixed states \cite{Namkung2017PRA}.

An interesting topic is to study
the relationship between different tasks in quantum information.
In this paper, we prove that SSD and LOCC
can be interpreted in a unified view,
despite the
essential distinction between the two protocols: the classical
communication is forbidden in the former one but required in the latter.

Different from SSD, another two
protocols called reproducing and discrimination after broadcasting, which allow classical communications, were discussed in Refs. \cite{Bergou2013PRL,Namkung2017PRA,Namkung2018SR} and compared
with the result of SSD.
It has been found that SSD performs better than the other two. It would be interesting to consider the compatibility of LOCC and these three protocols (SSD, reproducing and discrimination after broadcasting)
for our bipartite systems. In order to see the effects of different information tasks on the gap between
local and global schemes, we hybridize the LOCC protocol with these three protocols.
We show that the optimal successful probability
of global SSD can be attained by local SSD for some special cases. In contrast,
the local scheme is inferior to the global one for the other two protocols.

The paper is organized as follows. In Sec. \ref{mixed local state
discrimination}, we present the result of locally discriminating
mixed bipartite states, which is found to be equivalent to one
of the global schemes. In Sec. \ref{Simulation}, we simulate mixed-state discrimination by entangled pure states via local and global
schemes, respectively. In Sec. \ref{SSD},  we present a unified view
of SSD and LOCC protocols. In Sec. \ref{other
scenario}, we study the hybridization of LOCC and the other three
protocols. We summarize in the last section.

\section{mixed-state discrimination via local operations}\label{mixed local state discrimination}

\begin{figure}
\includegraphics[width=8cm]{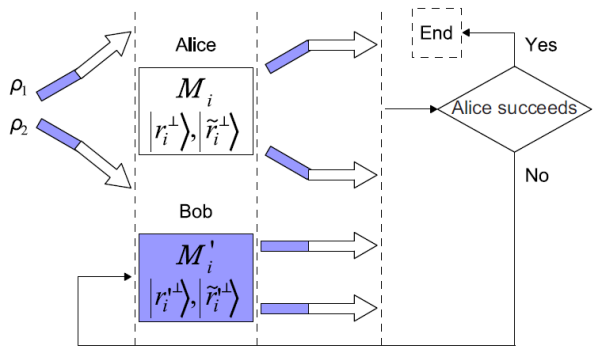} \\
 \caption{Protocol for local mixed-state discrimination. First, a
bipartite mixed quantum state
 $\rho_i$ ($i=1,2$)
 prepared with the prior probability $P_i$ is sent to
 Alice. Alice performs unambiguous discrimination on the state via
optimal local POVMs $\{M_i\}$ ($i=0,\ 1,\ 2$) on the subspace
 spanned by the basis $\{|r_i\rangle,|\tilde{r}_i\rangle\}$. Then, if Alice succeeds in
 discriminating the states optimally, the
 procedure ends; otherwise the state is sent to the other observer,
 Bob, who will discriminate the state optimally by the POVMs on the other subspace
 spanned by  $\{|r'_i\rangle,|\tilde{r}'_i\rangle\}$.} \label{fig1}
\end{figure}

We first consider the procedure of mixed-state discrimination via
local operations, (see Fig. \ref{fig1}). One
prepares an ensemble of two mixed bipartite separable states
$\rho_i$ with \emph{a priori} probability $P_i$, ($i=1,2$,
$P_1+P_2=1$). The state $\rho_i$ has a spectral decomposition,
\begin{equation}\label{initial state1}
\rho_i=r_i|r_i\rangle \langle r_i| \otimes|r_i'\rangle \langle
r_i'|+\tilde{r}_i|\tilde{r}_i\rangle \langle \tilde{r}_i|
\otimes|\tilde{r}_i'\rangle \langle \tilde{r}_i'|,~i=1,2
\end{equation}
with $r_i,\ \tilde{r}_i\in[0,1]$, $r_i+\tilde{r}_i=1$.
This mixed state is a statistical mixture of two vectors $|r_i\rangle\otimes|r_i'\rangle$ and $|\tilde{r}_i\rangle\otimes|\tilde{r}_i'\rangle$ with the classical probability $r_i$ and $\tilde{r}_i$, respectively.
We show that the existing results in Refs. \cite{Chen2001PRA} also hold for these bipartite
mixed states.
The vectors fulfill the following relations,
\begin{eqnarray}\label{overlap}
\langle r_1|r_2\rangle &=& s,~ \langle
 \tilde{r}_1|\tilde{r}_2\rangle=\tilde{s},~  \langle
 r'_1|r'_2\rangle=s',~
 \langle\tilde{r}'_1|\tilde{r}'_2\rangle=\tilde{s}',\nonumber\\
 \langle r_i|\tilde{r}_i\rangle &=& \langle r'_i|\tilde{r}'_i\rangle=0,
\end{eqnarray}
where $0<s,\ s',\ \tilde{s},\ \tilde{s}'<1$.
We assume that the support space of the two states do not overlap, namely,
\begin{equation}\label{zero overlap}
\langle r_1|\tilde{r}_2\rangle=\langle\tilde{r}_1|r_2\rangle=\langle
r'_1|\tilde{r}'_2\rangle=\langle\tilde{r}'_1|r'_2\rangle=0.
\end{equation}

The unambiguous state discrimination of two general mixed states with arbitrary support space is hard to be handled and solved analytically \cite{Namkung2017PRA}. Assumptions (\ref{overlap}) and (\ref{zero overlap}) ensure that the mixed states are not entangled and have no global coherence.
In the next section, we consider the role of entanglement (global coherence) in discriminating pure states superposed via the vectors $|r_i\rangle\otimes|r_i'\rangle$ and $|\tilde{r}_i\rangle\otimes|\tilde{r}_i'\rangle$.

Here, in each round, one of the two bipartite states (1) is sent to two observers, Alice and Bob. The first
(second) particle of the bipartite system, with local orthonormal
basis $\{|r_i\rangle,|\tilde{r}_i\rangle\}$
($\{|r'_i\rangle,|\tilde{r}'_i\rangle\}$), is provided for Alice
(Bob). Alice performs a local operation on the first particle with POVM
operators
\begin{eqnarray}\label{POVM}
&M^A_1&=c_1|r_2^{\bot}\rangle\langle
r_2^{\bot}|+\tilde{c}_1|\tilde{r}_2^{\bot}\rangle\langle\tilde{r}_2^{\bot}|,
\nonumber \\
&M^A_2&=c_2|r_1^{\bot}\rangle\langle
r_1^{\bot}|+\tilde{c}_2|\tilde{r}_1^{\bot}\rangle\langle\tilde{r}_1^{\bot}|,
\nonumber \\
&M^A_0&=I^A-M^A_1-M^A_2,
\end{eqnarray}
where \{$|r_1^{\bot}\rangle,\,|\tilde{r}_1^{\bot}\rangle$\} and
\{$|r_2^{\bot}\rangle,\,|\tilde{r}_2^{\bot}\rangle$\} are bases
orthogonal to \{$|r_1\rangle,\,|\tilde{r}_1\rangle$\} and
\{$|r_2\rangle,\,|\tilde{r}_2\rangle$\}, respectively. $c_i$ and $\tilde{c}_i$
are non-negative real numbers less than $1$.

Alice's POVM (\ref{POVM}) must satisfy the following three properties: (a) $M^A_i\geq 0$,
($i=0,1,2$), (b) $M^A_1+M^A_2+M^A_0=I^A$, and (c) ${\rm{Tr}}[\rho_i(M^A_j\otimes I^B)]=0$ for
$i,j=1,2$ and $i\neq j$. The first two relations are the positive-semidefinite
and completeness conditions of the POVM, respectively. The last one guarantees
no error occurring in state discrimination. The operators $M^A_1$
and $M^A_2$ are for the conclusive outcomes (corresponding to the
succeeding results) while $M^A_0$ is for the inconclusive one
(failure result).

From assumptions (\ref{overlap}) and (\ref{zero overlap}), one knows that the subspace spanned by $\{|r_1^{\bot}\rangle,\ |r_2^{\bot}\rangle\}$ is orthogonal to the one spanned by
$\{|\tilde{r}_1^{\bot}\rangle,\ |\tilde{r}_2^{\bot}\rangle\}$. Thus, the POVM in Eq. (\ref{POVM}) contains the direct sum ``$\oplus$" and Eq. (\ref{POVM}) can be written as
\begin{widetext}
\begin{eqnarray}\label{POVM direct sum}
&M^A_1&=c_1|r_2^{\bot}\rangle\langle
r_2^{\bot}|\oplus\tilde{c}_1|\tilde{r}_2^{\bot}\rangle\langle\tilde{r}_2^{\bot}|,
\nonumber \\
&M^A_2&=c_2|r_1^{\bot}\rangle\langle
r_1^{\bot}|\oplus\tilde{c}_2|\tilde{r}_1^{\bot}\rangle\langle\tilde{r}_1^{\bot}|,
\nonumber \\
&M^A_0&=M_{0s}^A\oplus\tilde{M}_{0s}^A=(I_s^A-c_1|r_1^{\bot}\rangle\langle
r_1^{\bot}|-c_2|r_2^{\bot}\rangle\langle
r_2^{\bot}|)
\oplus(\tilde{I}_s^A-\tilde{c}_1|\tilde{r}_1^{\bot}\rangle\langle\tilde{r}_1^{\bot}|
-\tilde{c}_2|\tilde{r}_2^{\bot}\rangle\langle\tilde{r}_2^{\bot}|).
\end{eqnarray}
\end{widetext}
Here, $I_s^A$ and $\tilde{I}_s^A$ are the identity matrices on their respective subspaces. The POVM with the form of Eqs. (\ref{POVM direct sum}) guarantees that the discrimination of mixed states can be carried out in two independent subspaces.

Set
$$
q^A_i=\langle r_i|M^A_{0s}|r_i\rangle,~~~\tilde{q}^A_i=\langle\tilde{r}_i|\tilde{M}^A_{0s}|\tilde{r}_i\rangle.
$$
We have
\begin{equation}\label{POVM parameter 1}
q_i^A=1-c_i(1-s^2),~~~
\tilde{q}_i^A=1-\tilde{c}_i(1-\tilde{s}^2),
\end{equation}
where $s^2\leq q_i^A\leq1, \tilde{s}^2\leq\tilde{q}_i^A\leq1$ ($i=1,2$).

As a positive-semidefinite operator, the POVM operator
$M^A_0$ satisfies ${\rm{det}} M^A_0\geq0$ \cite{Namkung2017PRA},
requiring that
\begin{eqnarray}\label{Alice measure}
q_1^Aq_2^A-s^2\geq0,~~~\tilde{q}_1^A\tilde{q}_2^A-\tilde{s}^2\geq0.
\end{eqnarray}
If all information encoded in the first particle of the bipartite
state is extracted by Alice, relations (\ref{Alice
measure}) become equalities. Otherwise, they are strict
inequalities.

The Kraus operators $K^A_i$ ($i=0,1,2$) corresponding to Alice's
POVM ($M^A_i=K_i^{A\dagger}K_i^A$) are given by \cite{Namkung2017PRA}
\begin{widetext}
\begin{equation}\label{Kraus operator}
\begin{aligned}
&K^A_1=\sqrt{c_1}|v_1\rangle\langle
r_2^{\bot}|+\sqrt{\tilde{c}_1}|\tilde{v}_1\rangle\langle
\tilde{r}_2^{\bot}|,& \\
&K^A_2=\sqrt{c_2}|v_2\rangle\langle
r_1^{\bot}|+\sqrt{\tilde{c}_2}|\tilde{v}_2\rangle\langle
\tilde{r}_1^{\bot}|,& \\
&K^A_0=(\sqrt{a_1}|v_1\rangle\langle
r_2^{\bot}|+\sqrt{a_2}|v_2\rangle\langle
r_1^{\bot}|)\oplus(\sqrt{\tilde{a}_1}|\tilde{v}_1\rangle\langle
\tilde{r}_2^{\bot}|+\sqrt{\tilde{a}_2}|\tilde{v}_2\rangle\langle
\tilde{r}_1^{\bot}|),&
\end{aligned}
\end{equation}
\end{widetext}
where $a_i=q_i^A/(1-s^2)$,
$\tilde{a}_i=\tilde{q}_i^A/(1-\tilde{s}^2)$, $i=1,2$.

The postmeasured state $\sigma_i$ ($i=1,2$) corresponding to
Alice's failure result can be expressed as
\begin{eqnarray}\label{postmeasurement state1}
\sigma_i &=& \frac{(K^A_0\otimes I^B)\rho_i(K_0^{A\dagger}\otimes
I^B)}{{\rm{Tr}}[(K^A_0\otimes I^B)\rho_i(K_0^{A\dagger}\otimes I^B)]}\nonumber\\[2mm]
&=& v_i|v_i\rangle\langle v_i|\otimes|r_i'\rangle \langle
r_i'|+\tilde{v}_i|\tilde{v}_i\rangle\langle\tilde{v}_i|\otimes|\tilde{r}_i'\rangle
\langle \tilde{r}_i'|.
\end{eqnarray}
Here, \{$|v_i\rangle,\,|\tilde{v}_i\rangle$\} is the orthonormal
basis of $\sigma_i$ satisfying
\begin{equation}\label{postmeasurement state overlap}
\begin{aligned}
&\langle v_1|v_2\rangle=t,\
\langle\tilde{v}_1|\tilde{v}_2\rangle=\tilde{t},\ \langle v_1|\tilde{v}_2\rangle=\langle\tilde{v}_1|v_2\rangle=0, &  \\
&\langle
v'_1|\tilde{v}'_2\rangle=\langle\tilde{v}'_1|v'_2\rangle=0.&
\end{aligned}
\end{equation}
$v_i$ and $\tilde{v}_i$ are the eigenvalues of $\sigma_i$,
\begin{eqnarray}\label{state parameter1}
v_i=\frac{q_i^Ar_i}{q_i^Ar_i+\tilde{q}_i^A\tilde{r}_i},~~~
\tilde{v}_i=\frac{\tilde{q}_i^A\tilde{r}_i}{q_i^Ar_i+\tilde{q}_i^A\tilde{r}_i}.
\end{eqnarray}
The equivalence of the two expressions (\ref{POVM direct sum}) and (\ref{Kraus operator}) ($M_0^A=K_0^{A\dagger}K_0^A$)
requires
\begin{equation}\label{POVM constraints}
q_1^Aq_2^A=s^2/t^2,\
\tilde{q}_1^A\tilde{q}_2^A=\tilde{s}^2/\tilde{t}^2.
\end{equation}
Then, one has that $q_i^A$ ($\tilde{q}_i^A$) is lower bounded by
$s^2/t^2$ ($\tilde{s}^2/\tilde{t}^2$).

The success probability for Alice to identify her state is
\begin{eqnarray}\label{PSSD}
P^{A}&=&\sum\limits_{i=1}^2P_i{\rm{Tr}}[\rho_i(M_i^A\otimes I^B)]\nonumber\\
&=&1-\sum\limits_{i=1}^2(P_ir_iq_i^A+P_i\tilde{r}_i\tilde{q}_i^A).
\end{eqnarray}
According to Eqs. (\ref{Alice measure}) and (\ref{POVM constraints}),
the optimal discrimination occurs when $t=\tilde{t}=1$.
Hence, relation (\ref{POVM constraints}) turns out to be
\begin{equation}\label{optimal POVM constraints}
q_1^Aq_2^A=s^2,\ \tilde{q}_1^A\tilde{q}_2^A=\tilde{s}^2
\end{equation}
with the parameters $q_i^{A}$ and $\tilde{q}_i^A$ satisfying
\begin{equation}\label{POVM constraints11}
q_i^{A}\in[s^2,1],\ \tilde{q}_i^{A}\in[\tilde{s}^2,1].
\end{equation}

The failure probability of Alice is easily acquired as
\begin{equation}\label{Alice failing1}
P^{A(f)}=1-P^{A}=P_1Q_1^A+P_2Q_2^A,
\end{equation}
where $Q_i^A=r_iq_i^A+\tilde{r}_i\tilde{q}_i^A$ ($i=1,2$).

If Alice succeeds in discriminating her state, the procedure ends.
Otherwise, another observer, Bob, performs unambiguous
discrimination via optimal local POVMs on the second particle
lying in the subspace spanned by the basis
$\{|r'_i\rangle,\,|\tilde{r}'_i\rangle\}$. Then, the \emph{a priori}
probability of Bob's states is
\begin{equation}\label{conditional prior1}
P_{fi}=\frac{P_iQ_i^A}{P_1Q_1^A+P_2Q_2^A},~~i=1,2.
\end{equation}
Construction of the optimal POVMs for Bob is similar to the ones for
Alice in Eqs. (\ref{POVM})-(\ref{POVM constraints11}) with the parameters $c_i$ and $\tilde{c}_i$ replaced by $c_i'$ and $\tilde{c}_i'$.  $q_i^A$ and
$\tilde{q}_i^A$ are also replaced by $q_i^B$ and
$\tilde{q}_i^B$, respectively, where
$$
q_i^B=1-c_i'(1-s'^2),~~~\tilde{q}_i^B=1-\tilde{c}_i'(1-\tilde{s}'^2).
$$
We also have the constraints
\begin{equation}\label{The constraints of Bob}
q_1^Bq_2^B=s'^2,\ \tilde{q}_1^B\tilde{q}_2^B=\tilde{s}'^2,
\end{equation}
with $q_i^B\in[s'^2,1]$ and $\tilde{q}_i^B\in[\tilde{s}'^2,1]$.

The success probability for Bob to identify his state is
\begin{eqnarray}\label{Bob succeeding1}
P^B&=&\sum\limits_{i=1}^2P_{fi}{\rm{Tr}}[\sigma_i(I^A\otimes M_i^B)] \nonumber\\
&=&\sum\limits_{i=1}^2P_{fi}[c_i'v_i(1-s'^2)+\tilde{c}'_i\tilde{v}_i(1-\tilde{s}'^2)] \nonumber\\
&=&\sum\limits_{i=1}^2P_{fi}[v_i(1-q_i^B)+\tilde{v}_i(1-\tilde{q}_i^B)].
\end{eqnarray}

Then, the failure probability for Bob's discrimination is given by
\begin{equation}\label{Bob failing1}
P^{B(f)}=1-P^B=P_{f1}Q_1^B+P_{f2}Q_2^B,
\end{equation}
where $Q_i^B=v_iq_i^B+\tilde{v}_i\tilde{q}_i^B$ ($i=1,2$).
Thus, we obtain the total failure probability
\begin{equation}\label{local failing of pure state}
P^{A(f)}P^{B(f)}=P_1Q_1^AQ_1^B+P_2Q_2^AQ_2^B.
\end{equation}

According to relations (\ref{optimal POVM constraints}),
(\ref{The constraints of Bob}), and (\ref{local failing of pure
state}), the total successful probability of the LOCC protocol is
\begin{eqnarray}\label{Total failing1}
P_{L}&=&1-P^{A(f)}P^{B(f)} \nonumber\\
&=&1-\sum\limits_{i=1}^2(P_ir_iq_i^L+P_i\tilde{r}_i\tilde{q}_i^L),
\end{eqnarray}
where $q_i^L=q_i^Aq_i^B$,
$\tilde{q}_i^L=\tilde{q}_i^A\tilde{q}_i^B$ ($i=1, 2$), and
$q_1^Lq_2^L=s_0^2$, $\tilde{q}_1^L\tilde{q}_2^L=\tilde{s}_0^2$
with $s_0=ss'$ and $\tilde{s}_0=\tilde{s}\tilde{s}'$. It can be
easily found that the result of Eq. (\ref{local failing of pure
state}) is invariant under exchanging Alice and Bob.

The success probability for the global scheme is given by
\begin{equation}\label{global failing of pure state}
P_{G}=1-\sum\limits_{i=1}^2(P_ir_iq_i^G+P_i\tilde{r}_i\tilde{q}_i^G),
\end{equation}
where the parameters $q_i^G$, $\tilde{q}_i^G$ ($i=1,2$)
satisfy the following relations:
\begin{equation}\label{The constraints of total one1}
q_1^Gq_2^G=s_0^2,\ \tilde{q}_1^G\tilde{q}_2^G=\tilde{s}_0^2.
\end{equation}
The result in Eq. (\ref{global failing of pure state}) is of the same form as the one for Alice's local scheme in
Eq. ($\ref{PSSD}$) with the parameters $q_i^A$, $\tilde{q}_i^A$, $s$, and
$\tilde{s}$ replaced by $q_i^G$, $\tilde{q}_i^G$, $s_0$, and
$\tilde{s}_0$, respectively.

Comparing the result of Eq. (\ref{Total failing1}) with Eq. (\ref{global
failing of pure state}), it is obvious that the local scheme is
equivalent to the global one. Setting $r_i=1$, we derive the pure state
$|\Psi_i\rangle=|r_i\rangle\otimes|r'_i\rangle$ ($i=1,2$). The overlap
$\langle\Psi_1|\Psi_2\rangle$ can be divided into two parties $s$
and $s'$. The relation $\langle\Psi_1|\Psi_2\rangle=ss'$ implies that the
difficulty for extracting information in local state
discrimination (two-step procedure) is identical to that in the global scheme
(one-step procedure). This condition guarantees the equivalence of
these two schemes. Relations (\ref{Total failing1}) and
(\ref{global failing of pure state}) indicate that the
successful probability for discrimination of a mixed state $\rho$ in
Eq. (\ref{initial state1}) is equivalent to a weighted average of the
one for two pairs of pure states $|r_1\rangle\otimes|r'_1\rangle$,
$|r_2\rangle\otimes|r'_2\rangle$ and
$|\tilde{r}_1\rangle\otimes|\tilde{r}'_1\rangle$,
$|\tilde{r}_2\rangle\otimes|\tilde{r}'_2\rangle$, lying in their
respective subspaces that are orthogonal to each other.  That is
a key reason why equivalence of local and global schemes
also occurs for the mixed states.

The optimal success probability of global mixed-state
discrimination, perfectly achieved by the local one, is shown in Table
\ref{tab} for $P_1r_1\leq P_2r_2$ and $P_1\tilde{r}_1\leq
P_2\tilde{r}_2$. The results are divided into four categories.  For the both-states-identified case in the
bottom right corner of Table \ref{tab}, we have
$P^{(opt)}=1-2\sqrt{P_1P_2}F(\rho_1,\rho_2)$ where
$F(\rho_1,\rho_2)=\sqrt{r_1r_2}ss'+\sqrt{\tilde{r}_1\tilde{r}_2}\tilde{s}\tilde{s}'$
is the fidelity \cite{Josza1994} between $\rho_1$ and $\rho_2$.
Here, the fidelity can be used as a generalized ``inner product" to
characterize the discrimination of mixed states. It can also be
easily concluded that as one mixed state is neglected in this
optimal solution (the case in the top left corner of Table
\ref{tab}), the optimal successful probability is $P_2(1-s_0^2)$ which is independent of $r_i$ ($\tilde{r}_i$) for $s_0=\tilde{s}_0$.

\begin{table*}
\begin{center}
\begin{tabular}{c c c c}
\hline \hline
$\textrm{Overlap}$                                            &$s_0>\sqrt{\frac{P_1r_1}{P_2r_2}}$                                 & $s_0\leq\sqrt{\frac{P_1r_1}{P_2r_2}}$      \\ \hline
& & \\
$\tilde{s}_0>\sqrt{\frac{P_1\tilde{r}_1}{P_2\tilde{r}_2}}$    &$P^{\max}=1-P_1-P_2(r_2s_0^2+\tilde{r}_2\tilde{s}_0^2)$    &$P^{\max}=1-2\sqrt{P_1r_1P_2r_2}s_0-P_1\tilde{r}_1-P_2\tilde{r}_2\tilde{s}_0^2$     \\
& & \\
$\tilde{s}_0\leq\sqrt{\frac{P_1\tilde{r}_1}{P_2\tilde{r}_2}}$ &~~~$P^{\max}=1-P_1r_1-P_2r_2s_0^2-2\sqrt{P_1\tilde{r}_1P_2\tilde{r}_2}\tilde{s}_0$~~~ &$P^{\max}=1-2\sqrt{P_1r_1P_2r_2}s_0-2\sqrt{P_1\tilde{r}_1P_2\tilde{r}_2}\tilde{s}_0$             \\ \hline \hline
\end{tabular}
\caption[table1]{\label{tab} Optimal success probability $P^{\max}$ of global mixed state discrimination in terms of $P_i$, $r_i$, $\tilde{r}_i$, $s_0$ and $\tilde{s}_0$. For $s_0\leq\sqrt{\frac{P_1r_1}{P_2r_2}}$, $\tilde{s}_0\leq\sqrt{\frac{P_1\tilde{r}_1}{P_2\tilde{r}_2}}$ , we have $q_1^G=\sqrt{\frac{P_2r_2}{P_1r_1}}s_0$ ($\tilde{q}_1^G=\sqrt{\frac{P_2\tilde{r}_2}{P_1\tilde{r}_1}}\tilde{s}_0$) while $q_1^G=1$ ($\tilde{q}_1^G=1$) for $s_0>\sqrt{\frac{P_1r_1}{P_2r_2}}$ ($\tilde{s}_0>\sqrt{\frac{P_1\tilde{r}_1}{P_2\tilde{r}_2}}$)  corresponding to this optimal
 solution. In the former case, both of the mixed states are identified;  while the latter case is considered to be one-state-identified because the success probability for identifying both $|r_1\rangle\langle r_1|\otimes|r_1'\rangle\langle r_1'|$ and $|\tilde{r}_1\rangle\langle \tilde{r}_1|\otimes|\tilde{r}_1'\rangle\langle \tilde{r}_1'|$ equals to zero
\cite{Zhang2017arXiv,Namkung2017PRA,Namkung2018SR}. If only one of
$|r_1\rangle\langle r_1|\otimes|r_1'\rangle\langle r_1'|$ and $|\tilde{r}_1\rangle\langle \tilde{r}_1|\otimes|\tilde{r}_1'\rangle\langle \tilde{r}_1'|$ is neglected and the other one is identified, we say
that the mixed state $\rho_1$ is partially identified (e.g. the
case for $s_0\leq\sqrt{\frac{P_1r_1}{P_2r_2}}$,
$\tilde{s}_0>\sqrt{\frac{P_1\tilde{r}_1}{P_2\tilde{r}_2}}$ with
the optimal success probability achieved at
$q_1^G=\sqrt{\frac{P_2r_2}{P_1r_1}}s_0$, $\tilde{q}_1^G=1$).}
\label{t2}
\end{center}
\end{table*}

From the above results one sees that conditions (\ref{overlap}) and (\ref{zero overlap}) greatly simplify the discrimination of mixed states (\ref{initial state1}).
Nevertheless, the POVM for more general cases, in which the constraints in Eqs. (\ref{overlap}) and (\ref{zero overlap}) on Bob's state are relaxed, can still be constructed.

In more general cases, the vectors $|r'_1\rangle$,
$|r'_2\rangle$, $|\tilde{r}'_1\rangle$, and $|\tilde{r}'_2\rangle$
overlap with each other. In order to discriminate the states unambiguously, Bob's POVM can be constructed in the form of Eq. (\ref{POVM}),
\begin{eqnarray}\label{POVM11}
&M_i^{B*}&=c'_i|\alpha_i\rangle\langle\alpha_i|+\tilde{c}'_i|\tilde{\alpha}_i\rangle\langle\tilde{\alpha}_i|,\nonumber\\
&M_0^{B*}&=I-M_1^{B*}-M_2^{B*},
\end{eqnarray}
where $c'_i$ and $\tilde{c}'_i$ ($i=1,2$) are also
non-negative parameters which satisfy $0<c'_i,~\tilde{c}'_i<1$. The vectors
$|\alpha_i\rangle$ and $|\tilde{\alpha}_i\rangle$ can be acquired
as \cite{Namkung2018SR},
\begin{eqnarray}\label{POVM vector}
&|\alpha_i\rangle&=\frac{\sum\limits_{j=1}^2G^{-1}_{2j-1,2i-1}|r_j'\rangle+G^{-1}_{2j,2i-1}|\tilde{r}_j'\rangle}
{||\sum\limits_{j=1}^2G^{-1}_{2j-1,2i-1}|r_j'\rangle+G^{-1}_{2j,2i-1}|\tilde{r}_j'\rangle||},\nonumber\\
&|\tilde{\alpha}_i\rangle&=\frac{\sum\limits_{j=1}^2G^{-1}_{2j-1,2i}|r_j'\rangle+G^{-1}_{2j,2i}|\tilde{r}_j'\rangle}
{||\sum\limits_{j=1}^2G^{-1}_{2j-1,2i}|r_j'\rangle+G^{-1}_{2j,2i}|\tilde{r}_j'\rangle||},
\end{eqnarray}
where $G$ is the Gram matrix \cite{Namkung2017PRA,Namkung2018SR} given by the four
vectors $|r_i'\rangle$, $|\tilde{r}_i'\rangle$ ($i=1,2$):
\begin{eqnarray}\label{Gram matrix}
 G=\left [
 \begin{matrix}
   \langle r_1'|r_1'\rangle & \langle r_1'|\tilde{r}_1'\rangle& \langle r_1'|r_2'\rangle &\langle r_1'|\tilde{r}_2'\rangle  \\
   \langle \tilde{r}_1'|r_1'\rangle & \langle \tilde{r}_1'|\tilde{r}_1'\rangle& \langle \tilde{r}_1'|r_2'\rangle &\langle \tilde{r}_1'|\tilde{r}_2'\rangle  \\
   \langle r_2'|r_1'\rangle & \langle r_2'|\tilde{r}_1'\rangle& \langle r_2'|r_2'\rangle &\langle  r_2'|\tilde{r}_2'\rangle  \\
   \langle \tilde{r}_2'|r_1'\rangle & \langle \tilde{r}_2'|\tilde{r}_1'\rangle& \langle \tilde{r}_2'|r_2'\rangle &\langle \tilde{r}_2'|\tilde{r}_2'\rangle
  \end{matrix}
  \right  ].
\end{eqnarray}
It is easily verified that
${\rm{Tr}}[\sigma_i(I^A\otimes M^B_j)]=0$ ($i,\ j=1,2, i\neq j$).
Thus, Bob's success probability can also be acquired
according to Eq. (\ref{Bob succeeding1}). Nevertheless, because the
subspace spanned by \{$|\alpha_1\rangle$, $|\alpha_2\rangle$\} is
not orthogonal to the one spanned by \{$|\tilde{\alpha}_1\rangle$,
$|\tilde{\alpha}_2\rangle$\} anymore, it is difficult to optimize the success probability. We have the following conjecture.

\emph{[Conjecture 1].}
For a fixed fidelity between $\rho_1$ and $\rho_2$, the success probability
for discriminating the mixed states is impaired by the overlaps of the vectors
$|r_i'\rangle$ and $|\tilde{r}_j'\rangle$ ($i,j=1,2$).

Namely, the local scheme is inferior to the global one. However, it is difficult to prove the general case for the conjecture, as Bob's success probability has a complex form. In Appendix \ref{mixed non-orthogonal state}, we give a proof for a special case in which the overlaps are the same as in Eqs. (2) and (3), with only one of the zero overlaps replaced by $\langle r_2'|\tilde{r}_2'\rangle=\varepsilon$.

\section{Simulation for mixed state discrimination}\label{Simulation}

In the above results, the vectors are mixed via classical probabilities $r_i$ and $\tilde{r}_i$
($i = 1, 2$).
In this part, we turn to study the case in which they are coherently superposed into the pure states.
Namely, we are simulating mixed-state discrimination by identifying a pair of pure entangled (coherent) states of the form
\begin{equation}\label{initial pure state}
|\Psi_i\rangle=\sqrt{r}_i|r_i\rangle \otimes|r_i'\rangle
+\sqrt{\tilde{r}_i}|\tilde{r}_i\rangle \otimes|\tilde{r}_i'\rangle,~~~i=1,2
\end{equation}
occurring with the \emph{a priori} probability $P_i$.
The parameters $r_i$ and $\tilde{r}_i$  are neither $1$
nor zero. $|\Psi_i\rangle$ are both entangled and globally coherent,
with the same fidelity as the mixed states (\ref{initial state1}) shown as
\begin{equation}\label{condition of fidelity}
F(|\Psi_1\rangle,|\Psi_2\rangle)=F(\rho_1,\rho_2)=\sqrt{r_1r_2}ss'+\sqrt{\tilde{r}_1\tilde{r}_2}\tilde{s}\tilde{s}'.
\end{equation}

For the local scheme, we adopt the same protocol as the one in Fig. \ref{fig1}, and the local POVMs
$M_i$ and Kraus operators $K_i$ ($i=0,1,2$) used by
Alice and Bob given in Eqs. (\ref{POVM}) and (\ref{Kraus operator}); we have Alice's failure probability
\begin{eqnarray}
P_E^{A(f)}&=&P_1\langle\Psi_1|M_0^A\otimes I^B|\Psi_1\rangle+P_2\langle\Psi_2|M_0^A\otimes I^B|\Psi_2\rangle\nonumber\\
&=&P_1r_1q_1^A+P_1\tilde{r}_1\tilde{q}^A_1+P_2r_2q^A_2+P_2\tilde{r}_2\tilde{q}^A_2.
\end{eqnarray}
Corresponding to Alice's failure result, the postmeasured state for Bob is given by
\begin{equation}\label{post-measured state}
\frac{K_0\otimes I|\Psi_i\rangle}{||K_0\otimes
I|\Psi_i\rangle||}=\sqrt{v_i}|v\rangle\otimes |r'_i\rangle
+\sqrt{\tilde{v}_i}|\tilde{v}\rangle\otimes |\tilde{r}'_i\rangle,
\end{equation}
occurring with the \emph{a priori} probability
\begin{equation}
P_i^0\!=\!\frac{P_ir_iq_i^A+P_i\tilde{r}_i\tilde{q}_i^A}{P_1r_1q_1^A\!+\!P_1\tilde{r}_1\tilde{q}_1^A\!
+\!P_2r_2q_2^A\!+\!P_2\tilde{r}_2\tilde{q}_2^A}~~~i=1,2,
\end{equation}
where $v_i$ and $\tilde{v}_i$ are given in Eq. (\ref{state
parameter1}).

Bob's failure probability
\begin{equation}
P_E^{B(f)}=P_1^0v_1q^B_1+P_1^0\tilde{v}_1\tilde{q}^B_1+P_2^0v_2q^B_2+P_2^0\tilde{v}_2\tilde{q}^B_2,
\end{equation}
is the same as Eq. (\ref{Bob failing1}). The
total failure probability $P_E^{A(f)}P_E^{B(f)}$ is identical to
$P^{A(f)}P^{B(f)}$ in Eq. (\ref{local failing of pure state}).
Consequently, this simulation is perfect for the local scheme. But for
the global scheme, we have a completely different conclusion.

\emph{[Theorem 1].} For the global scheme, the optimal success probability of discriminating
the pure entangled states [Eq. (\ref{initial pure
state})] is achieved perfectly by the result of
mixed states for the both-states-identified case,
$s_0\leq\sqrt{\frac{P_1r_1}{P_2r_2}}$,
$\tilde{s}_0\leq\sqrt{\frac{P_1\tilde{r}_1}{P_2\tilde{r}_2}}$,
and for the one-state-identified case,
$s_0>\sqrt{\frac{P_1r_1}{P_2r_2}}$,
$\tilde{s}_0>\sqrt{\frac{P_1\tilde{r}_1}{P_2\tilde{r}_2}}$ under the condition
$\sqrt{r_1\tilde{r}_2}\tilde{s}_0=\sqrt{r_2\tilde{r}_1}s_0$.
Otherwise, the results are superior to the ones of mixed states.

\emph{[Proof].} The global scheme is to discriminate a pair of
nonorthogonal states $|\Psi_1\rangle$ and $|\Psi_2\rangle$ with an
inner product $s^*=\langle
\Psi_1|\Psi_2\rangle=\sqrt{r_1r_2}s_0+\sqrt{\tilde{r}_1\tilde{r}_2}\tilde{s}_0$.
This comes down to an optimization problem:
\begin{equation}
{\rm{maximize}}\ P_{SUCC}=1-P_1q_1^0-P_2q_2^0,
\end{equation}
\begin{equation}\label{inner product}
{\rm{subject\ to}}\ \,q_1^0q_2^0=s^{*2}, ~q_1^0,q_2^0\in[s^{*2},\ 1].
\end{equation}
We have the optimized success probability
\begin{subequations}\label{Succeeding of pure state}
\begin{align}
\mathrm{(i):\ }&P_{(E)G}^{\max}=1-2\sqrt{P_1P_2}s^* &{\rm{when}}\ s^*\leq\sqrt{\frac{P_1}{P_2}},   \\
\mathrm{(ii):\ }&P_{(E)G}^{\max}=P_2(1-s^{*2}) &{\rm{when}}\
s^*>\sqrt{\frac{P_1}{P_2}}.
\end{align}
\end{subequations}

\begin{figure}
\includegraphics[width=8cm]{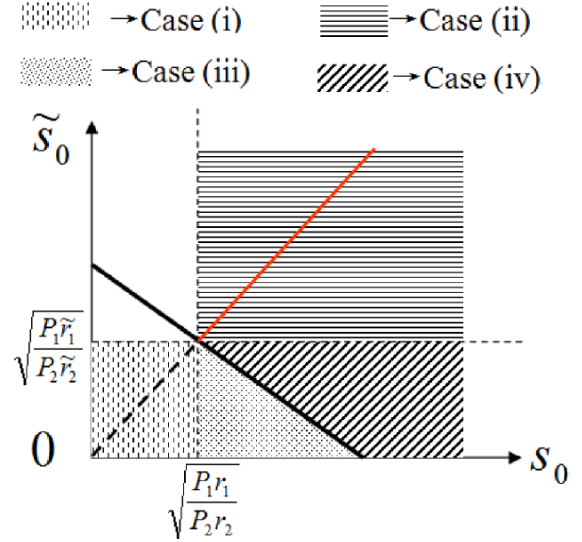} \\
 \caption{Four regions corresponding to cases (i), (ii), (iii), and
(iv), respectively, with different values of $s_0$, $\tilde{s}_0$,
and $s^*$. The black solid line stands for
$s^*=\sqrt{r_1r_2}s_0+\sqrt{\tilde{r}_1\tilde{r}_2}\tilde{s}_0=\sqrt{\frac{P_1}{P_2}}$
with fixed $r_i$ and $\tilde{r}_i$ ($i=1,\ 2$). The red solid line
stands for $\sqrt{r_1\tilde{r}_2}\tilde{s}_0=\sqrt{r_2\tilde{r}_1}s_0$.} \label{fig2}
\end{figure}

Let us compare this result with the one of mixed-state
discrimination in Table \ref{tab}. The four cases corresponding
to different value ranges of $s_0$, $\tilde{s}_0$, and $s^*$ are
shown in Fig. \ref{fig2}.

Case (i) ($s_0\leq\sqrt{\frac{P_1r_1}{P_2r_2}}$,
$\tilde{s}_0\leq\sqrt{\frac{P_1\tilde{r}_1}{P_2\tilde{r}_2}}$). We
can easily obtain $s^*\leq\sqrt{P_1/P_2}$. Here, both the mixed
and pure entangled states are all optimally identified. The
optimized successful probability for discriminating mixed states
is given by
\begin{eqnarray}\label{result of case (i)}
P^{\max}&=&1-2\sqrt{P_1r_1P_2r_2}s-2\sqrt{P_1\tilde{r}_1P_2\tilde{r}_2}\tilde{s}\nonumber\\
&=&1-2\sqrt{P_1P_2}s^*=P_{(E)G}^{\max}.
\end{eqnarray}

Case (ii) ($s_0>\sqrt{\frac{P_1r_1}{P_2r_2}}$,
$\tilde{s}_0>\sqrt{\frac{P_1\tilde{r}_1}{P_2\tilde{r}_2}}$). We
can also easily acquire  $s^*>\sqrt{\frac{P_1}{P_2}}$. The optimal
discrimination of both pure-entangled and mixed states is
the one-state-identified case. According to the result in
Eq. (\ref{Succeeding of pure state}b) and the Cauchy-Schwarz inequality,
one has
\begin{eqnarray}
P_{(E)G}^{\max}&=&P_2(1-s^{*2})\nonumber\\
&=&P_2[1-(\sqrt{r_1r_2}s_0+\sqrt{\tilde{r}_1\tilde{r}_2}\tilde{s}_0)^2]\nonumber\\
&\geq&P_2\{1-(r_1+\tilde{r}_1)(r_2s_0^2+\tilde{r}_2\tilde{s}_0^2)\}\nonumber\\
&=&1\!-\!P_1\!-\!P_2(r_2s_0^2\!+\!\tilde{r}_2\tilde{s}_0^2)\!=\!P^{\max}.
\end{eqnarray}
When $\sqrt{r_1\tilde{r}_2}\tilde{s}_0=\sqrt{r_2\tilde{r}_1}s_0$
(red solid line in Fig. \ref{fig2}), the relation becomes an
equality.

Case (iii) ($s_0>\sqrt{\frac{P_1r_1}{P_2r_2}}$,
$\tilde{s}_0\leq\sqrt{\frac{P_1\tilde{r}_1}{P_2\tilde{r}_2}}$,
$s^*\leq\sqrt{\frac{P_1}{P_2}}$). The optimal discrimination of
pure-entangled (mixed) states is the both-state-identified
(one-state-partially-identified) case. We have
\begin{eqnarray}
\Delta P&=&P_{(E)G}^{\max}-P^{\max}\nonumber\\
&=&(1-2\sqrt{P_1P_2}s^*)\nonumber\\
& &-(1-P_1r_1-P_2r_2s_0^2-2\sqrt{P_1\tilde{r}_1P_2\tilde{r}_2}\tilde{s}_0)\nonumber\\
&=&(\sqrt{P_1r_1}-\sqrt{P_2r_2}s_0)^2>0.
\end{eqnarray}

Case (iv) ($s_0>\sqrt{\frac{P_1r_1}{P_2r_2}}$,
$\tilde{s}_0\leq\sqrt{\frac{P_1\tilde{r}_1}{P_2\tilde{r}_2}}$,
$s^*>\sqrt{\frac{P_1}{P_2}}$). The optimal discrimination of
pure-entangled (mixed) states is the one-state-identified
(one-state-partially-identified) case. We have
\begin{eqnarray}\label{Partial-state-identified case 1}
\Delta P&=&P_{(E)G}^{\max}-P^{\max}\nonumber\\
&=&P_2(1-s^{*2})\nonumber\\
& &-(1-P_1r_1-P_2r_2s_0^2-2\sqrt{P_1\tilde{r}_1P_2\tilde{r}_2}\tilde{s}_0)\nonumber\\
&=&F(\tilde{s}_0)=A\tilde{s}_0^2+B\tilde{s}_0+C,
\end{eqnarray}
where
\begin{equation}\label{Partial-state-identified case 2}
\begin{aligned}
&A=-P_1(1-r_1)(1-r_2),&\\
&B=2\sqrt{(1\!-\!r_1)(1\!-\!r_2)}(\sqrt{P_1P_2}\!-\!P_2\!-\!P_2\sqrt{r_1r_2}s_0),&\\
&C=-(1-r_1)(P_1-P_2r_2s_0^2).&
\end{aligned}
\end{equation}

The $\Delta P$ given in Eq. (\ref{Partial-state-identified case 1}) can be viewed as
a quadratic function of the variable $\tilde{s}_0$
with $\tilde{s}_0\in(0,\sqrt{\frac{P_1\tilde{r}_1}{P_2\tilde{r}_2}}]$.
Because of $A<0$, the minimum of $\Delta P$ is obtained at the
boundary points, $\Delta
P_{\rm{min}}={\rm{min}}\{\Delta P|_{\tilde{s}_0\to0},\ \Delta
P|_{\tilde{s}_0=\sqrt{\frac{P_1\tilde{r}_1}{P_2\tilde{r}_2}}}\}$.

According to the constraint $s^*>\sqrt{\frac{P_1}{P_2}}$ with
$\tilde{s}_0\to0$, we have
\begin{equation}
\sqrt{r_1r_2}s_0>\sqrt{\frac{P_1}{P_2}}.
\end{equation}
Then, we can easily get
\begin{eqnarray}\label{boundary point 1}
\Delta P|_{\tilde{s}_0\to0}&=&-(1-r_1)(P_1-P_2r_2s_0^2)\nonumber\\
&>&-(1-r_1)(P_1-P_2r_2\frac{P_1}{r_1r_2P_2})\nonumber\\
&=&\frac{P_1(1-r_1)^2}{r_1}>0.
\end{eqnarray}
For another boundary point
$\tilde{s}_0=\sqrt{\frac{P_1\tilde{r}_1}{P_2\tilde{r}_2}}$,
according to Eqs. (\ref{Partial-state-identified case 1}) and
(\ref{Partial-state-identified case 2}), we have
\begin{equation}
\Delta
P|_{\tilde{s}_0=\sqrt{\frac{P_1\tilde{r}_1}{P_2\tilde{r}_2}}}
=(1-r_1)(\sqrt{P_2r_2}s_0-\sqrt{P_1r_1})^2>0.
\end{equation}

Due to the symmetry of exchanging $s_0$ and
$\tilde{s}_0$, another one-state-partially-identified case,
$s_0\leq\sqrt{\frac{P_1r_1}{P_2r_2}}$,
$\tilde{s}_0>\sqrt{\frac{P_1\tilde{r}_1}{P_2\tilde{r}_2}}$, for
the optimal discrimination of mixed states leads to the same
conclusions as cases (iii) and (iv). $\Box$

\begin{figure}
\includegraphics[width=8cm]{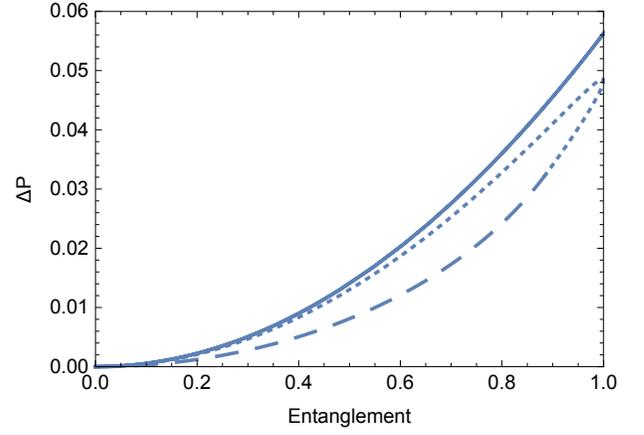} \\
 \caption{The difference of the optimal success probability $\Delta P$ between the two schemes as functions of the entanglement $E(|\Psi_i\rangle)$ between the two particles corresponding to the cases for $P_1=0.1$, $P_2=0.9$, $s_0=0.7$, and $\tilde{s}_0=0.2$. Solid line, case (ii); dotted line, case (iii); dashed line, case (iv).} \label{fig3}
\end{figure}

Obviously, the difference between the optimal success
probability for globally discriminating the pure and mixed states
equals the one between the local and global schemes for the pure-entangled-state protocol itself.
This difference arises from the coherent superposition of bipartite vectors.
It also indicates that entanglement plays a key role in the process of global discrimination scheme for the pure states (\ref{initial pure state}).
Then, let us consider the difference $\Delta P=P_{(E)G}^{\max}-P^{\max}$ as a function of the entanglement $E(|\Psi_i\rangle)$ between the two
particles. Set $r_1=r_2=r$. Based on the negativity entanglement
measure \cite{Lee2000PRL}, we have $E(|\Psi_i\rangle)=2\sqrt{r(1-r)}$.
From Fig. \ref{fig3} one can see that the difference $\Delta P$
increases with the entanglement (as well as the global coherence),
which is more obvious for the case (ii)
(one-mixed-state-identified case). Consequently, for global pure state discrimination, entanglement
between the two particles is a kind of critical recourse
which is completely destroyed by local operations.
Thus, the pure-entangled-state protocol cannot reflect any superiority versus
mixed-state one in the local scheme. The effect of entanglement
(global coherence) vanishes for special cases mentioned in
\emph{Theorem 1} where successful simulation occurs.

From the above results, it is indicated that relation (\ref{condition of fidelity})
is a necessary condition for successful simulation.
If this condition is violated, the results differs.
For example, suppose that the pure entangled state in (\ref{initial pure state}) is changed into
\begin{equation}\label{initial pure state11}
|\psi_i\rangle\!=\!\sqrt{r}_i|r_i\rangle\!\otimes\!|r_i'\rangle
\!+\!\exp{(i\phi_i)}\sqrt{\tilde{r}_i}|\tilde{r}_i\rangle
\!\otimes\!|\tilde{r}_i'\rangle,~i=1,2,
\end{equation}
with $\phi_i$ as a phase factor satisfying $\phi_2\neq
\phi_1+2k\pi$ for some integer $k$; for the global scheme we have
\begin{eqnarray}
F(|\Psi_1\rangle,|\Psi_2\rangle)&=&\left|\langle\psi_1|\psi_2\rangle\right|\nonumber\\
&=&\left|\sqrt{r_1r_2}s_0+\exp{[i(\phi_2-\phi_1)]}\sqrt{\tilde{r}_1\tilde{r}_2}\tilde{s}_0\right|\nonumber\\
&<&\left|\sqrt{r_1r_2}s_0\right|+\left|\sqrt{\tilde{r}_1\tilde{r}_2}\tilde{s}_0\right|\nonumber\\
&=&\sqrt{r_1r_2}s_0+\sqrt{\tilde{r}_1\tilde{r}_2}\tilde{s}_0\nonumber\\
&=&s^*=F(\rho_1,\rho_2).
\end{eqnarray}
Here, the optimal success probability in discriminating the states is
bound to be superior to the result in Eq. (\ref{Succeeding of pure
state}). That is, even for case (i), this simulation fails.

\section{Unified view of SSD and LOCC protocol}\label{SSD}

\begin{figure}
\includegraphics[width=8cm]{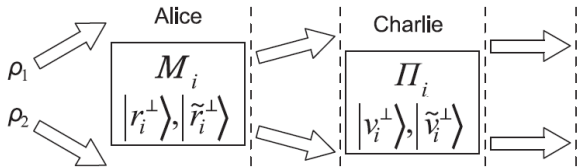} \\
 \caption{SSD protocol presented in Refs.
\cite{Bergou2013PRL,Namkung2017PRA}. First, a mixed quantum
state
 $\rho_i$ ($i=1,2$)
 prepared with the \emph{a priori} probability $P_i$ is sent to
 Alice. Alice performs unambiguous discrimination on the state via
 nonoptimal global POVMs $\{M_i\}$ ($i=0,\ 1,\ 2$). Then, Alice sends her
 postmeasured state $\sigma_i$ to Charlie and Charlie
 identifies $\sigma_i$ with an optimal POVM on the same particle.
 The classical communications between Alice and Charlie are forbidden in this procedure.} \label{fig4}
\end{figure}

The protocol of SSD mentioned in Ref. \cite{Bergou2003PRL} has been
extended to mixed initial states in \cite{Namkung2017PRA}.
We suppose that one prepares a mixed
state (\ref{initial state1}) and sends
it to Alice (see Fig. \ref{fig4}). Alice's POVMs and postmeasured states are of the
same form as Eqs. (\ref{POVM}) and (\ref{postmeasurement state1}).
Here, Alice's POVMs are nonoptimal, meaning that $t\neq1$ and
$\tilde{t}\neq 1$. Namely, after Alice's discrimination, there is
some information left in her state. Then, the postmeasured state
is sent to another observer, Charlie. Charlie will discriminate the
state via POVMs on the same particle,
different from the LOCC protocol in Sec. \ref{mixed local state
discrimination}. The POVMs are given by
\begin{eqnarray}
\Pi_1&=&\frac{1-q^C_1}{1-t^2}|v_2^{\bot}\rangle\langle
v_2^{\bot}|+\frac{1-\tilde{q}^C_1}{1-\tilde{t}^2}|\tilde{v}_2^{\bot}\rangle\langle\tilde{v}_2^{\bot}|,
\nonumber \\
\Pi_2&=&\frac{1-q^C_2}{1-t^2}|v_1^{\bot}\rangle\langle
v_1^{\bot}|+\frac{1-\tilde{q}^C_2}{1-\tilde{t}^2}|\tilde{v}_1^{\bot}\rangle\langle\tilde{v}_1^{\bot}|,
\nonumber \\
\Pi_0&=&I-\Pi_1-\Pi_2,
\end{eqnarray}
where \{$|v_1^{\bot}\rangle$, $|\tilde{v}_1^{\bot}\rangle$\} and
\{$|v_2^{\bot}\rangle$, $|\tilde{v}_2^{\bot}\rangle$\} are bases
orthogonal to \{$|v_1\rangle$, $|\tilde{v}_1\rangle$\} and
\{$|v_2\rangle$, $|\tilde{v}_2\rangle$\}, respectively. Here,
$q^C_i=\langle v_i|\Pi_0|v_i\rangle$, $\tilde{q}^C_i=\langle
\tilde{v}_i|\Pi_0|\tilde{v}_i\rangle$, $i=1,2$.

Charlie's discrimination is optimal, in the sense that
${\rm{det}} \Pi_0=0$, i.e., $q_1^Cq_2^C-t^2=0$ and
$\tilde{q}_1^C\tilde{q}_2^C-\tilde{t}^2=0$. The joint success
probability for both Alice and Charlie to identify the states is
\begin{eqnarray}\label{PSSD1}
\!\!P_{SSD}^{A(s),C(s)}\!\!&\!\!=\!\!\!\!&\sum\limits_{i=1}^2P_i{\rm{Tr}}[\rho_i
M_i]{\rm{Tr}}[\sigma_i\Pi_i]\nonumber\\
&\!\!=\!\!\!&\sum\limits_{i=1}^2P_i\![r_i\!(1\!\!-\!\!q_i^A)\!
(1\!\!-\!\!q_i^C)\!\!+\!\!\tilde{r}_i\!(1\!\!-\!\!\tilde{q}_i^A)\!(1\!\!-\!\!\tilde{q}_i^C\!)\!].
\end{eqnarray}
Its optimization of $P_{SSD}^{A(s),C(s)}$ has been given in
Ref. \cite{Namkung2017PRA}.

During the procedure of SSD, classical communications are
forbidden
\cite{Bergou2013PRL,Pang2013PRA,Namkung2017PRA,Zhang2017arXiv}.
This is essentially different from the local scheme where Bob's
discrimination of the second particle is dependent on the premise
that Alice communicates her failure result to him.
The outcomes about Alice's succeeding, Bob's failing, or both succeeding are rejected by the LOCC scheme. Despite this distinction, we can show that the SSD and local protocol
can be interpreted in a unified way: The information Alice and Charlie extract in the process of SSD is equivalent to that encoded in the first and second particle in LOCC which is distributed to Alice and Bob respectively. Then, we have the following theorem.

\emph{[Theorem 2].} If the POVMs used by the
observer Bob (in local scheme) and Charlie (in the SSD scheme) satisfy
$q_i^B=q_i^C$ and $\tilde{q}_i^B=\tilde{q}_i^C$ ($i=1,2$), the
probability that at least one of Alice and Charlie succeeds
\cite{Bergou2013PRL,Zhang2017arXiv} in SSD is equal to the total
succeeding probability of the LOCC protocol.

\emph{[Proof].} For the SSD protocol, the probability that at least one
of Alice and Charlie succeeds in detecting the state is given by
\begin{equation}\label{one succeeds}
P_{SSD}^{A,C(1)}=P_{SSD}^{A(f),C(s)}+P_{SSD}^{A(s),C(f)}+P_{SSD}^{A(s),C(s)},
\end{equation}
which includes three parts $P_{SSD}^{A(f),C(s)}$, $P_{SSD}^{A(s),C(f)}$,
and $P_{SSD}^{A(s),C(s)}$, standing for the probability that Alice
fails (succeeds), Bob succeeds (fails), and both succeed,
respectively. By straightforward calculations, we have
\begin{equation}\label{One of Alice or Bob fails}
\begin{aligned}
&P_{SSD}^{A(f),C(s)}=\sum\limits_{i=1}^2[P_ir_iq_i^A(1-q_i^C)+P_i\tilde{r}_i\tilde{q}_i^A(1-\tilde{q}_i^C)],\\
&P_{SSD}^{A(s),C(f)}=\sum\limits_{i=1}^2[P_ir_i(1-q_i^A)q_i^C+P_i\tilde{r}_i(1-\tilde{q}_i^A)\tilde{q}_i^C].
\end{aligned}
\end{equation}
Combining Eqs. (\ref{PSSD1})-(\ref{One of Alice or Bob fails}), one has
\begin{eqnarray}\label{One succeeds1}
P_{SSD}^{A,C(1)}&=&\sum\limits_{i=1}^2[P_ir_i(1-q_i^Aq_i^C)+P_i\tilde{r}_i(1-\tilde{q}_i^A\tilde{q}_i^C)]\nonumber\\
&=&1-\sum\limits_{i=1}^2(P_ir_iq_i^Aq_i^C+P_i\tilde{r}_i\tilde{q}_i^A\tilde{q}_i^C),
\end{eqnarray} which is equal to $P_L$ in Eq. (\ref{Total failing1}) for
$q_i^B=q_i^C$. $\Box$

\section{Hybridization between LOCC and other scenarios}\label{other scenario}

In Sec. \ref{Simulation}, it is indicated that coherent superposition of the bipartite vectors
leads to the difference between local and global schemes in state discrimination. In this section, we hybridize LOCC with three protocols
(SSD, reproducing, and discrimination after broadcasting) in order to see whether different information tasks (e.g. sequential observation or classical communications between different observers) contribute to this gap.  The latter two scenarios are introduced in Refs. \cite{Bergou2013PRL,Namkung2017PRA} to compare with SSD in order to see the effect of classical communications on state discrimination.

(1) Reproducing protocol: The observer Alice performs an optimal unambiguous discrimination
 measurement on the quantum state $\rho_i$ prepared with the probability $P_i$. If she
succeeds, she reproduces the state and sends it to Charlie;
if she fails, she informs Charlie that his measurement
failed, and that is the end of the procedure.

(2) Discrimination after broadcasting: Broadcasting \cite{Li2009JPA,Namkung2017PRA} is identical to the quantum cloning \cite{Duan1998} for the
pure-state case. It transforms a mixed state $\rho$ into
$\rho_{AC}$ satisfying ${\rm{Tr}}_{_A}\rho_{AC}={\rm{Tr}}_{_C}\rho_{AC}=\rho$ with a
certain success probability.  If Alice succeeds in broadcasting, she shares the state $\rho_{AC}$ with
Charlie and they all perform optimal POVM on the partial states.  If the broadcasting fails,
she informs Charlie, and that is the end of the procedure.

The results in Refs. \cite{Bergou2013PRL,Namkung2017PRA} indicate that SSD performs better than these two strategies.
In order to perform our hybridizations, four
observers Alice, Bob, Charlie, and David will cooperate to
discriminate the bipartite state (\ref{initial state1}) in three ways. The first
(second) particle of the bipartite system is provided for Alice
and Charlie (Bob and David).
\begin{figure}
\includegraphics[width=8cm]{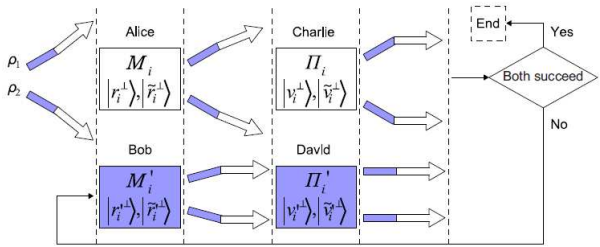} \\
 \caption{Protocol for SSD via local operations. A bipartite
quantum state $\rho_i$, $i=1,2$, prepared with the prior
probability $P_i$, is sent to Alice. Alice and Charlie perform
the SSD procedure on the first particle lying in the subspace spanned
by the basis $\{|r_i\rangle,|\tilde{r}_i\rangle\}$ to discriminate
the state. If both of them succeed, the procedure ends.
Otherwise, the post-measured state is sent to other two observers,
Bob and David, who perform another SSD procedure on the other
particle to discriminate the state by their POVMs on the other
subspace spanned by $\{|r'_i\rangle,|\tilde{r}'_i\rangle\}$.}
\label{fig5}
\end{figure}

(i) Hybridization of LOCC with SSD (see Fig. \ref{fig5}):  Although classical
communications are forbidden in the process of SSD, we suppose
that Alice and Charlie are allowed to check their results with
each other after they finish their measurements. If both of them
succeed, the procedure ends. Otherwise, Bob and
David perform another SSD procedure on the second
particle of the bipartite system.

(ii) Hybridizing LOCC with protocols (1) and (2): Alice and Charlie perform
the reproducing (discrimination after broadcasting) protocol on the first particle of our
bipartite state $\rho_i$. If both of them succeed, the procedure ends. Otherwise, Bob and
David perform another one on the second particle.

We enumerate examples for discriminating bipartite pure
states (\ref{initial state1}) with
$r_1=r_2=1$. The difference
of optimal successful probabilities $\Delta P$ between global and
local SSD has been calculated in detail (see Appendixes B and C). For the
hybridization of LOCC with reproducing protocol, the calculations
of the optimal POVM is the same as the one in Sec. \ref{mixed
local state discrimination}. The \emph{a priori} probability of the
states left for Bob and David is equal to $1/2$. Then, the
difference of optimal success probability between the global and
local schemes is given by
\begin{eqnarray}
\Delta P^{(Re)}&=&P^{opt}_{Re,G}-P^{opt}_{Re,L}\nonumber\\
&=&P^{opt(f)}_{Re,L}-P^{opt(f)}_{Re,G}\nonumber\\
&=&\![1\!-\!(1\!-\!s)^2][1\!-\!(1\!-\!s')^2]\!-\![1\!-\!(1\!-\!ss')^2]\!\nonumber\\
&=&2ss'(1-s)(1-s')>0.
\end{eqnarray}

The successful probability for broadcasting two equal-prior pure states with
the inner product $s$ is $1/(1+s)$ \cite{Bergou2013PRL,Namkung2017PRA,Duan1998}.  Then, for the hybridization of LOCC with discrimination after broadcasting, the \emph{a priori} probability of the state
left for Bob and David is $1/2$ as well. We also obtain the difference
of optimal success probability between the global and local schemes:
\begin{eqnarray}
\Delta P^{(Br)}&=&P^{opt}_{Br,G}-P^{opt}_{Br,L} \nonumber\\
&=&\![1\!-\!\frac{(1\!-\!s)^2}{1+s}]\![1\!-\!\frac{(1-s')^2}{1+s'}]\!
-\![1\!-\!\frac{(1-ss')^2}{1+ss'}]\!\nonumber\\
&=&\frac{2(1-s)(1-s')ss'(3+ss')}{(1+s)(1+s')(1+ss')}>0.
\end{eqnarray}

\begin{figure}
\includegraphics[width=8cm]{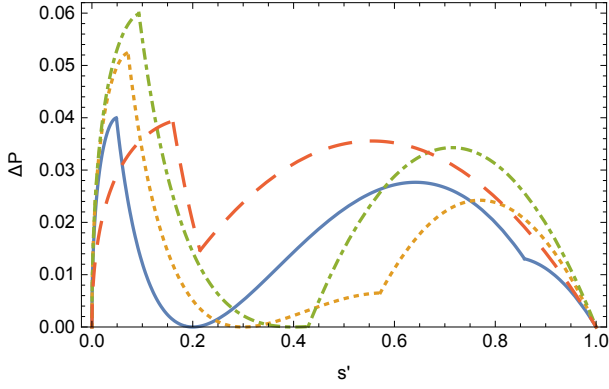} \\
 \caption{The difference of the optimal success probability $\Delta P$ between global and local SSD vs $s'$.
Solid line: $s=0.2$; dotted line: $s=0.3$; dot-dashed line:
$s=0.4$; dashed line: $s=0.8$. If $s=s'$, the
difference vanishes for a special value ranges of $s$ and $s'$ (e.g. $s=0.2,\ 0.3,\ 0.4$),
as shown in Appendix \ref{Protocol1b}. Otherwise, this difference is
positive (e.g. $s=0.8$).} \label{fig6}
\end{figure}

It is seen that using hybridization of LOCC with the other three
protocols in which classical communication occurs to guarantee more
observers to succeed extends the gap between the optimal success
probability of the global and local schemes. We prove that for special
cases, the result of global SSD can be achieved by the local one. In
contrast, the local scheme is inferior to the global one for the other
two protocols. Some of these results are given in Fig.
\ref{fig6}.

\section{Summary and Outlook}\label{Summ}

We have extended the local discrimination of bipartite pure states \cite{Chen2001PRA} to
rank-2 mixed ones \cite{Namkung2017PRA} via a statistical mixture of two pairs of state
vectors. Assuming that these two vectors are orthogonal to each other and the support space of the two mixed states does not overlap, we have shown that the local scheme can perform as well as the scheme with global measurements, just as the result for pure initial states \cite{Chen2001PRA}.
An example shows that the local scheme is inferior to the global one
if this condition is not satisfied.

Then, the mixed (separable) state discrimination is simulated by pure entangled
states, with the factors of classical probability in mixed states
replaced by quantum probability amplitudes in pure states. It has been shown that
this simulation is perfect for the local scheme because
local POVM eliminates the entanglement and global coherence
encoded in the pure entangled state. Thus, the
pure-entangled-state protocol does not show any superiority to
the mixed one. For the global
scheme, successful simulation only occurs for a few
special cases.

For the global scheme, this perfect simulation also
occurs when the following two conditions are satisfied: (i) the
fidelity of the mixed states equals that of the pure entangled
states and (ii) both of the mixed states are identified. Otherwise, except for a few
special one-state-identified cases, the mixed-state protocol is
inferior to the pure-entangled-state one.

Concerning another SSD protocol given in
Refs. \cite{Bergou2013PRL,Namkung2017PRA,Namkung2018SR} which is useful in quantum communication schemes (e.g., the B92 quantum cryptography protocol \cite{Bennett1992PRL}), we have obtained an interesting result: the
probability for at least one of the two observers succeeding in
SSD is equal to the total succeeding probability of the local schemes.
Thus, in spite of an essential distinction (classical
communication is forbidden in SSD but required in the local scheme)
between the two protocols, the SSD and LOCC protocols can be
interpreted in a unified way.

At last, after hybridizing LOCC with the other three protocols (SSD, reproducing and broadcasting),
we have found that the gap between the optimal success probabilities of the global and local schemes
is extended. The result of the global scheme can be achieved by the local
one only for SSD but not for the other two protocols.

We can easily get a generalized result for many-body systems.
The successful probability is equivalent to the result of SSD in
consecutive observers discussed partly in Ref. \cite{Hillery2017JPA} if
we require at least one observer succeeding. Namely, this unified
view for discrimination in $N$-body states and SSD in $N$
consecutive observers also holds. Simulation of $N$-body mixed
states by pure entangled states can also be similarly studied. Moreover,
by introducing an ancillary system coupled with the principal one
\cite{Zhang2013SR,Pang2013PRA,Zhang2017arXiv,Kim2018arXiv}, the
Hilbert space can be extended. Thus, a POVM can be realized via
the tensor product method \cite{Chen2007PRA}. The role of quantum
correlation \cite{Zhang2013SR,Pang2013PRA,Zhang2017arXiv} and
coherence \cite{Kim2018arXiv,Xiong2018JPA} in pure-state
discrimination has been also studied. It is also desirable to investigate
the requirement of quantum correlations and coherence in mixed-state discrimination procedures via the Neumark formalism.

\begin{acknowledgments}
This work is supported by NSF of China (Grant No. 11675119,
11675113, 11575125), Beijing Municipal Commission of Education (KZ201810028042),
and Beijing Natural Science Foundation (Z190005).
\end{acknowledgments}

\appendix
\section{A special case for Conjecture 1} \label{mixed non-orthogonal state}

We consider a special case for the discrimination of
mixed state (\ref{initial state1}) with a replaced condition
$\langle r'_2|\tilde{r}'_2\rangle=\varepsilon$ that does not affect the fidelity between $\rho_1$ and $\rho_2$. Assume that $|r'_1\rangle=|0\rangle$ and $|\tilde{r}'_1\rangle=|1\rangle$.
We have
\begin{eqnarray}\label{non-orthogonal base}
|r_2'\rangle\!&=&\!s'|0\rangle+\sqrt{1-s'^2}|2\rangle,\nonumber\\
|\tilde{r}_2'\rangle\!&=&\! \tilde{s}'|1\rangle\!\!+\!\!\frac{\varepsilon}{\sqrt{1\!\!-\!\!s'^2}}|2\rangle\!\!+\!\!\sqrt{1-\tilde{s}'^2\!-\!\frac{\varepsilon^2}{1\!-\!s'^2}}|3\rangle,
\end{eqnarray}
where $\varepsilon$ satisfies $0<\varepsilon\leq\sqrt{(1-s'^2)(1-\tilde{s}'^2)}$. According to Eq. (\ref{Gram matrix}), the Gram matrix can be written as
\begin{eqnarray}\label{Gram matrix1}
 G=\left [
 \begin{matrix}
   1 & 0& s' &0  \\
  0 & 1&  0 &\tilde{s}'  \\
   s' & 0&1 &\varepsilon  \\
   0 & \tilde{s}'& \varepsilon &1
  \end{matrix}
  \right  ].
\end{eqnarray}
Then, according to Eqs. (\ref{POVM11}), (\ref{POVM vector}),  (\ref{Gram matrix1}), and (\ref{Bob succeeding1}), Bob's success probability of discriminating the mixed states is given by
\begin{eqnarray}\label{Bob's success probability}
\!\!P^{B*}\!\!&\!\!=\!\!&P_{f1}{\rm{Tr}}[\sigma_1(I^A\otimes M_1^{B*})]+P_{f2}{\rm{Tr}}[\sigma_2(I^A\otimes M_2^{B*})], \nonumber \\
&\!\!=\!\!&T(\!\varepsilon\!)(\frac{P_{f1}v_1c_1'}{\!1\!-\!\tilde{s}'^2\!-\!\varepsilon^2\!}\!\!+\!\!\frac{\!P_{f1}\!\tilde{v}_1\!\tilde{c}_1'\!}{\!1\!-\!s'^2\!-\!\varepsilon^2\!}\!\!+\!\!\frac{\!P_{\!f\!2\!}v_2c_2'\!}{1-\tilde{s}'^2}\!\!+\!\!\frac{\!P_{\!f\!2\!}\tilde{v}_2\tilde{c}_2'\!}{1-s'^2}),
\end{eqnarray}
where we have set $T(\varepsilon)=(1-s'^2)(1-\tilde{s}'^2)-\varepsilon^2$.

Compared with the result in Eq. (\ref{Bob succeeding1}), we obtain  $P^{B*}<P^B$ and $\lim\limits_{\varepsilon\to0}P^{B*}=P^B$. The gap between $P^{B*}$ and $P^B$ still exists when the mixed states approach pure ones (e.g., $v_1,\,v_2\to1$). This fact shows the discontinuous points ($v_i=0,1$, $i=1,2$) of the success probability.

\section{Optimal SSD with both states identified by Alice and Charlie}\label{Protocol1a}
For two initial bipartite pure states with
$r_1=r_2=1$, shared by Alice and Charlie and prepared with equal priority,
we consider the optimal success probability of the local SSD protocol
and compare it with the global one for the both-state-identified case
of Alice and Charlie ($0<s\le 3-2\sqrt{2}$
\cite{Pang2013PRA,Namkung2017PRA}).

Based on Eq. (\ref{PSSD1}), for the local SSD protocol of pure states
prepared with equal priority ($P_1=P_2=1/2$), the optimization of the success
probability $P^{A,C}_{SSD,L}$ for Alice and Charlie is given by the following:
\begin{equation}\label{Alice and Charlie succeeds}
{\rm{maximize}}\
\!P\!^{A\!,C\!}_{\!SSD\!,\!L}\!=\!P_1(1\!-\!q_1^A)(1\!-\!q_1^C)\!+\!P_2\!(1\!-\!q_2^A)(1\!-\!q_2^C)\!,
\end{equation}
\begin{align}
\begin{aligned}
{\rm{subject\ to}}\ &q_1^Aq_2^C=\frac{s^2}{t^2}, \ q_1^A, q_2^A\in[\frac{s^2}{t^2},1], \\
&q_1^Cq_2^C=t^2,\ q_1^C,\ q_2^C\in[t^2,1].
\end{aligned}
\end{align}

The optimal success probability
\begin{equation}\label{Alice and Chalie's optimal SSD1}
P_{SSD}^{A,C\rm{(opt)}}=(1-\sqrt{s})^2
\end{equation}
occurs for
$q_1^{A}=q_1^{C}=q_2^{A}=q_2^{C}=\sqrt{s}$, $t=\sqrt{s}$,  for
both-state-identified case ($0<s\leq3-2\sqrt{2}$)
\cite{Pang2013PRA,Namkung2017PRA}. The \emph{a priori} probability
of Bob's states is shown as
\begin{equation}
P_{fi}=\frac{P_i[1-(1-q_i^A)(1-q_i^C)]}
{\sum\limits_{i=1}^2\{P_i[1-(1-q_i^A)(1-q_i^C)]\}}=1/2.
\end{equation}

In a similar way, the optimization of the success
probability of Bob and David's local SSD can be given as follows:
\begin{equation}
{\rm{maximize}}\
\!P^{\!B\!,D\!}_{\!SSD,\!L}=\!P_{\!f1}(1\!-\!q_1^B)(1\!-\!q_1^D)+\!P_{f2}\!(1\!-\!q_2^B)(1\!-\!q_2^D)\!,
\end{equation}
\begin{align}
\begin{aligned}
{\rm{subject\ to}}\ &q_1^Bq_2^B=\frac{s'^2}{t'^2}, \ q_1^B, q_2^B\in[\frac{s'^2}{t'^2},1], \\
&q_1^Dq_2^D=t'^2,\ q_1^D,\ q_2^D\in[t'^2,1].
\end{aligned}
\end{align}
Since the \emph{a priori} probability of the two states for Bob and
David is equal, the optimal successful probability of their SSD is \cite{Pang2013PRA}
\begin{subequations}\label{maxb}
\begin{align}
\mathrm{(i)\ \ \ }&\!P^{\!B,D\!\rm{(opt)}\!}_{SSD,L}\!=\!(1\!-\!\sqrt{s'})^2 \!\!&{\mathrm{when}}\ 0<s'\leq3-2\sqrt{2}, \\
\mathrm{(ii)\ \ \
}&\!P^{\!B,\!D\rm{(opt)}}_{SSD,L}\!=\!1/2(1\!-\!s')^2\! \!\!
&{\mathrm{when}}\ 3-2\sqrt{2}\!<\!s'\!<\!1.\!
\end{align}
\end{subequations}
For case (i), the optimal SSD occurs at $q_1^B=q_1^D=\sqrt{s'}$,
$t'=\sqrt{s'}$, whereas it occurs at $q_1^B=q_1^D=1$, $t'=\sqrt{s'}$ for case
(ii), where Bob and David conspire to ignore $\rho_1$. Then, the
total failure probability corresponding to this optimal local SSD
is
\begin{equation}\label{total failure SSD}
P^{\rm{opt}(f)}_{SSD,L}=(1-P^{A,C(\rm{opt})}_{SSD,L})(1-P^{B,D(\rm{opt})}_{SSD,L}).
\end{equation}

The optimal success probability of the global scheme is equivalent to the
result in Eqs. (\ref{maxb}) if we replace the inner product factors
$s'$ and $\tilde{s}'$ by $ss'$ and $\tilde{s}\tilde{s}'$,
respectively. The failure probability corresponding to the optimal
global SSD is
\begin{subequations}\label{global SSD}
\begin{align}
\mathrm{(i)\ \ \ }&P^{\rm{opt}(f)}_{SSD,G}=1-(1-\sqrt{ss'})^2, \nonumber\\
 &{\mathrm{when}}\ 0< ss'\leq 3-2\sqrt{2}; \\
\mathrm{(ii)\ \ \ }&P^{\rm{opt}(f)}_{SSD,G}=1-1/2(1-ss')^2, \nonumber\\
&{\mathrm{when}}\ 3-2\sqrt{2}<ss'< 1.
\end{align}
\end{subequations}

Nevertheless, here only case (i) in Eq. (\ref{global SSD}a) is possible.
Then, according to Eqs. (\ref{Alice and Chalie's optimal SSD1}), (\ref{maxb}), (\ref{total failure SSD}), and (\ref{global SSD}a), the difference in the optimal success probabilities between global
and local SSD ($\Delta P=P^{\rm{opt}(f)}_{SSD,L}-P^{\rm{opt}(f)}_{SSD,G}$) is given by
\begin{subequations}\label{maxc}
\begin{align}
\mathrm{(i)\ \ \  }&\Delta P=2\sqrt{ss'}(1-\sqrt{s'})(1-\sqrt{s}),  \\
\mathrm{(ii)\ \ \ }&\Delta P=\sqrt{s}(1-\sqrt{s'})F(s,s'), \\
\nonumber
\end{align}
\end{subequations}
with $F(s,s')=(2-\sqrt{s})(1+\sqrt{s'})(1+s')-4\sqrt{s'}$. Cases (i) and (ii) correspond to $0<s'\leq3-2\sqrt{2}$ and
$3-2\sqrt{2}<s'<1$, respectively. It can be easily acquired that
$\Delta P$ is bound to be positive for case (i). Since
$0<s\leq3-2\sqrt{2}$, for case (ii) we have $F(s,s')\geq
F(3-2\sqrt{2},s')$ and
\begin{equation}
\begin{aligned}
&\frac{dF(3-2\sqrt{2},s')}{ds'}|_{s'=s_0}=0,&\\
&\frac{d^2F(3-2\sqrt{2},s')}{ds'^2}|_{s'=s_0}\approx9.11>0&
\end{aligned}
\end{equation}
where $s_0=(29+12\sqrt{2}-2\sqrt{154+84\sqrt{2}})/63$. Hence, we get the minimum:
\begin{equation}
\min\limits_{s'}F(3-2\sqrt{2},s')=F(3-2\sqrt{2},s_0)\approx0.96>0.
\end{equation}
Therefore, we have $F(s,s')>0$ and $\Delta P>0$
according to relation (\ref{maxc}b).

\section{Optimal SSD with one-state-identified by Alice and Charlie}\label{Protocol1b}

For $3-2\sqrt{2}<s<1$, the optimization of the result in
Eq. (\ref{Alice and Charlie succeeds}) is achieved as
\begin{equation}\label{Alice and Chalie's optimal SSD}
P_{SSD}^{A,C(\rm{opt})}=1/2(1-s)^2
\end{equation}
for $q_1^{A}=q_1^{C}=1$, $q_2^{A}=q_2^{C}=s$, $t=\sqrt{s}$, where $\rho_1$ is conspired to
be ignored by Alice and Charlie \cite{Pang2013PRA,Namkung2017PRA}.

\begin{figure}
\includegraphics[width=8cm]{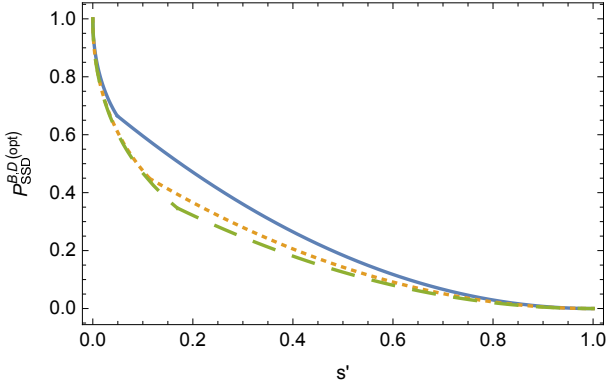} \\
 \caption{The joint optimal success probability of Bob and David as a function of $s'$. Solid line: $s=0.2$; dotted line: $s=0.5$; dashed line: $s=0.9$.} \label{fig7}
\end{figure}

We obtain the \emph{a priori} probabilities from the two
states left for Bob and David:
\begin{equation}
P_{f1}=\frac{\frac{1}{2}-\frac{1}{2}(1-s)^2}{1-\frac{1}{2}(1-s)^2},~~
P_{f2}=\frac{\frac{1}{2}}{1-\frac{1}{2}(1-s)^2}.
\end{equation}
Using a random search method \cite{Zhang2017arXiv}, we can seek
out the optimized success probability of SSD for both Bob and
David. For fixed $s$, we have $P_{f1}\leq1/2$.  The optimized success
probability occurs at $t'=\sqrt{s'}$ and $q_1^B=q_1^D$, which
indicates the equivalence between the information extracted by Bob
and David. The result of this optimization is given by
\begin{align}\label{max SSD of Bob and David}
\begin{aligned}
\mathrm{(i)} &~P_{SSD}^{B,D(\rm{opt})}=P_{f1}(1-q^*)^2+P_{f2}(1-\frac{s'}{q^*})^2,
\\ &~{\rm{when}}  \ 0<s' \le  s^c; \\
\mathrm{(ii)} &~P_{SSD}^{B,D(\rm{opt})}=P_{f2}(1-s')^2,\ {\rm{when}}\
s^c<s'<1,
\end{aligned}
\end{align}
where $q^*$ satisfies
$P_{f1}(q^*)^4-P_{f1}(q^*)^3+P_{f2}s'q^*-P_{f2}s'^2=0$, and the
critical value $s^c$  is determined by
$P_{f1}(1-q^*)^2+P_{f2}(1-{s^c}/{q^*})^2=P_{f2}(1-s^c)^2$. For
case (i), the optimal success probability occurs at
$q_1^B=q_1^D=q^*$, while it occurs at $q_1^B=q_1^D=1$ for case (ii), where Bob
and David conspire to ignore the state $\rho_1$. In Fig.
\ref{fig7}, it is shown that as $s$ decreases, Bob's state tends
to be equal prior. And the critical value $s^c$ approaches its maximum
$3-2\sqrt{2}$, which is consistent with the result in Ref. \cite{Zhang2017arXiv}.

\begin{figure}
\includegraphics[width=8cm]{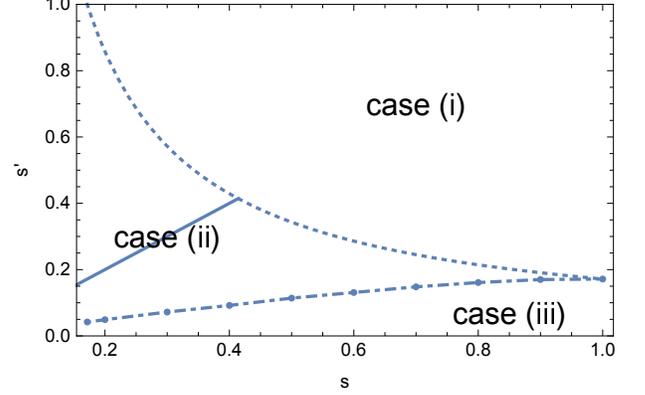} \\
 \caption{The dotted line ($ss'=3-2\sqrt{2}$) and dot-dashed line (a set for
$s'=s^c$ plotted numerically corresponds to different values of
$s$) are two bounds which give rise to three regions corresponding
to cases (i) $s^c<s'<1$, $3-2\sqrt{2}<ss'<1$, (ii) $s^c<s'<1$,
$0<ss'\leq 3-2\sqrt{2}$ and (iii) $0<s'\leq s^c$, $0<ss'\leq
3-2\sqrt{2}$, respectively. Only for case (ii) with $s=s'$ (solid
line), the optimal successful probability of local SSD attains the
result of global one.} \label{fig8}
\end{figure}

According to Eqs.(\ref{Alice and Chalie's optimal SSD}), (\ref{max SSD of Bob and David}), and (\ref{total failure SSD}), the total failure
probability of the optimal local SSD protocol can also be obtained.
For the global protocol, the failure probability of the optimal SSD can
be obtained from the result in Eqs. (\ref{global SSD})
with two possible outcomes. The difference of the optimal successful
probability between the global and local protocols is acquired
corresponding to the following three cases (i), (ii), and (iii),
as shown intuitively in Fig. \ref{fig8}.

\emph{Case (i)}: $s^c<s'<1$, $3-2\sqrt{2}<ss'<1$.  We have
\begin{eqnarray}\label{case (i)}
\Delta
P&=&\![1\!-\!\frac{1}{2}\!(1\!-\!s)^2][1\!-\!P_{f2}\!(1-\!s')^2]\!-\![1\!-\!\frac{1}{2}\!(1\!-\!ss')^2] \nonumber\\
&=&\frac{1}{2}(1-s)(1-s')(s+s'+ss'-1).
\end{eqnarray}
Since $s'>\frac{3-2\sqrt{2}}{s}$, we get
\begin{eqnarray}
s'-\frac{1-s}{1+s}&>&\frac{3-2\sqrt{2}}{s}-\frac{1-s}{1+s}\nonumber\\
&=&\frac{(s-\sqrt{2}+1)^2}{s(1+s)}\geq0,
\end{eqnarray}
from which we have $s+s'+ss'-1>0$. Hence, from Eq. (\ref{case
(i)}), it is easily obtained that $\Delta P>0$ as well.

\emph{Case (ii)}: $s^c<s'<1$, $0<ss'\leq 3-2\sqrt{2}$. We have
\begin{eqnarray}
\Delta
P&=&\![1\!-\!\frac{1}{2}(1\!-\!s)^2]\![1\!-\!P_{f2}(1\!-\!s')^2]\!-\![1\!-\!(1\!-\!\sqrt{ss'})^2]\! \nonumber\\
&=&\frac{1}{2}(\sqrt{s}-\sqrt{s'})^2[2-(\sqrt{s}+\sqrt{s'})^2]\geq0.
\end{eqnarray}
As $s=s'$, we get $\Delta P=0$. Namely, the optimal
success probability of the global SSD is attained by the local one.

\emph{Case (iii)}: $0<s'\leq s^c$, $0<ss'\leq3-2\sqrt{2}$. This is a
complicated case and is difficult to solve
analytically. By numerical experiment via $10^5$ random
numbers, it can be ensured that $\Delta P$ is also larger than zero.

\end{document}